\RequirePackage{fix-cm}
\documentclass[smallextended]{svjour3}       
\smartqed  
\usepackage{graphicx}
\usepackage[authoryear]{natbib}

\usepackage{graphicx}
\usepackage{fixltx2e}
\usepackage{mathtools}
\usepackage{ctable}
\usepackage{enumitem}

\begin{document}
\title{Patent Retrieval: A Literature Review}



\author{Walid Shalaby         \and
        Wlodek Zadrozny.
}


\institute{Walid Shalaby \at
              Department of Computer Science \\ 
              University of North Carolina at Charlotte \\
              9201 University City Blvd, Charlotte, NC 28223, USA \\
              \email{wshalaby@uncc.edu}           
           \and
           Wlodek Zadrozny \at
              Department of Computer Science \\ 
              University of North Carolina at Charlotte \\
              9201 University City Blvd, Charlotte, NC 28223, USA \\
              \email{wzadrozn@uncc.edu}           
}

\date{Received: 13 November 2017 / Accepted: 22 November 2018}

\maketitle

\begin{abstract}
	With the ever increasing number of filed patent applications every year, the need for effective and efficient systems for managing such tremendous amounts of data becomes inevitably important. Patent Retrieval (PR) is considered the pillar of almost all patent analysis tasks. PR is a subfield of Information Retrieval (IR) which is concerned with developing techniques and methods that effectively and efficiently retrieve relevant patent documents in response to a given search request. In this paper we present a comprehensive review on PR methods and approaches. It is clear that, recent successes and maturity in IR applications such as Web search cannot be transferred directly to PR without deliberate domain adaptation and customization. Furthermore, state-of-the-art performance in automatic PR is still around average in terms of recall. These observations motivate the need for interactive search tools which provide cognitive assistance to patent professionals with minimal effort. These tools must also be developed in hand with patent professionals considering their practices and expectations. We additionally touch on related tasks to PR such as patent valuation, litigation, licensing, and highlight potential opportunities and open directions for computational scientists in these domains.


\end{abstract}

\section{Introduction}
\label{intro}
Patents represent proxies for economic, technological, and even social activities. The Intellectual Property (IP) system motivates the disclosure of novel technologies and ideas by granting inventors exclusive monopoly rights on the economic value of their inventions. Patents, therefore, have a major impact on enterprises market value \cite{piroi2011clef}. With the continuous rise in the number of filed patent applications every year, the need for effective and efficient systems for managing such tremendous amounts of data becomes inevitably important. 

Typical patent analysis tasks include: 1) technology exploration in order to capture new and trendy technologies in a specific domain, and subsequently using them to create new innovative services, 2) technology landscape analysis in order to assess the density of patent filings of specific technology, and subsequently direct R\&D activities accordingly, 3) competitive analysis and benchmarking in order to identify strengths and differences of corporate's own patent portfolio compared to other key players working on related technologies, 4) patent ranking and scoring in order to quantify the strength of the claims of an existing or a new patent, and 5) prior art search in order to retrieve patent documents and other scientific publications relevant to a new patent application. All those patent-related activities require tremendous level of domain expertise which, even if available, must be integrated with highly sophisticated and intelligent analytics that provide cognitive and interactive assistance to the users.

Patent Retrieval (PR) is the pillar of almost all patent analysis tasks. PR is a subfield of Information Retrieval (IR) which is concerned with developing techniques and methods that effectively and efficiently retrieve relevant patent documents in response to a given search request. Although the field of IR has received huge advances from decades of research and development, research in PR is relatively newer and more challenging. On the one hand, patents are multi-page, multi-modal, multi-language, semi-structured, and metadata rich documents. On the other hand, patent queries can be a complete multi-page patent application. These unique features make traditional IR methods used for Web or ad hoc search inappropriate or at least of limited applicability in PR.

Moreover, patent data is multi-modal and heterogeneous. As indicated by \cite{lupu2011current}, analyzing such data is a challenging task for many reasons; patent documents are lengthy with highly complex and domain specific terminology. To establish their work novelty, inventors tend to use jargon and complex vocabulary to refer to the same concepts. They also use vague and abstract terms in order to broaden the scope of their patent protection making the problem of patent analysis linguistically challenging. 

PR starts with a search request (query) which often represents a patent application under novelty examination. Therefore, several methods for query reformulation (QRE) have been proposed in order to select, remove, or expand terms in the original query for improved retrieval. QRE methods are either keyword-based, semantic-based, or interactive. Keyword-based methods work by searching for exact matches between search query terms and the target corpus, and thus fail to retrieve relevant documents which use different vocabulary but have similar meaning to the reformulated query. In order to alleviate the vocabulary mismatch problem, semantic-based methods try to search by meaning through expanding queries and/or target corpus with similar or related terms and thus bridging the vocabulary gap. Because neither methods proved acceptable performance, few interactive methods were proposed to allow users to interactively control QRE with reasonable effort.

This review aims to provide researchers with an illustrative and critical overview of recent trends, challenges, and opportunities in PR. The rest of this paper is organized as follows. Section \ref{patent-retrieval-preliminaries} presents some preliminaries and background about patents data. Section \ref{patent-retrieval-data} provides an overview of evaluation tracks and data collections for PR benchmarking. An illustration of PR tasks is presented in Section \ref{patent-retrieval-tasks}. Section \ref{patent-retrieval-methods} presents a comprehensive review on PR methods and approaches. Section \ref{patent-retrieval-related}  lightly touches on related tasks such as patent quality assessment, litigation, and licensing. Finally, concluding remarks are presented in Section \ref{patent-retrieval-conclusion}.

\begin{table}
	{\renewcommand{\arraystretch}{1.4}
		\setlength{\tabcolsep}{1.8pt}
		\centering
		\caption{Patent Kind Codes of Major Patent Offices}
		\label{tbl:kind codes}
		\begin{tabular}{|c|l|l|l|}
			\hline
			\textbf{Type} & \multicolumn{1}{c|}{\textbf{USPTO (US)}} & \multicolumn{1}{c|}{\textbf{EPO (EP)}} & \multicolumn{1}{c|}{\textbf{WIPO (WO)}} \\ \hline
			A1            & application                         & \multicolumn{2}{c|} {application w/ search report}       \\ \hline
			A2            & republished application             & \multicolumn{2}{c|}{application w/o search report}      \\ \hline
			A3            & \multicolumn{1}{c|}{-}              & \multicolumn{2}{c|} {search report}                      \\ \hline
			A4            & \multicolumn{1}{c|}{-}              & supplementary search report       & publication of amended claims      \\ \hline
			A9            & \multicolumn{3}{c|}{modified application} \\ \hline
			B1            & granted patent w/o application               & granted patent (publication)      & \multicolumn{1}{c|}{-}             \\ \hline
			B2            & granted patent w/ application                & amended B1                        & \multicolumn{1}{c|}{-}             \\ \hline
		\end{tabular}
	}
\end{table}

\section{Preliminaries}
\label{patent-retrieval-preliminaries}
\subsection{Patent Documents and Kind Codes}
Patent documents are mostly textual. They are highly structured with typical elements (sections) including {\it title, abstract, background of the invention, description and claims}. The {\it description} section articulate in details the technical specification of the invention and its possible embodiments. The {\it claims} section is the most significant one as it describes the scope of protection sought by the inventor and hence encodes the real value of the patent. Patent documents are lengthy with highly complex and domain specific terminology. They also contain multiple data types (e.g., text, images, flow charts, formulae...etc) with a rich set of metadata and bibliographic information (e.g., {\it classification codes}, {\it citations}, {\it inventors}, {\it assignee}, {\it filing/publication dates}, {\it addresses}, {\it examiners}...etc). 

Typically, each patent has a set of pertaining documents which are published throughout its life-cycle. All documents are identified by an alphanumeric name with a common naming convention. Names start with two letters identifying the issuing patent office (e.g., US and EP), then the patent number as sequence of digits, and finally a suffix indicating the document's kind code. The kind code identifies the stage in the patent life-cycle at which the document is published. Table \ref{tbl:kind codes} shows a brief description of kind codes used at major patent offices and organizations including the US Patent and Trademark Office\footnote{http://www.uspto.gov/} (USPTO), the European Patent Office\footnote{http://www.epo.org/} (EPO) and the World Intellectual Property Organization\footnote{http://www.wipo.int/portal/en/index.html}(WIPO). 

\subsection{Patent Classification}
Patent offices organize patents by assigning classification codes to each of them based on the technical features of the invention. The patent classification system is a hierarchical one. Common classification systems include the International Patent Classification (IPC) and the Cooperative Patent Classification (CPC). The CPC was jointly developed by the USPTO and EPO to replace the US Patent Classification (USPC) and European classification system (ECLA). 
\subsection{Patent Families}
A patent family is a collection of patents that refer to the same invention and are granted at different countries around the world (\cite{piroi2013}). Typically, they describe a single invention in different languages depending on the issuing patent office. In the context of PR and prior art search, patents belonging to the same patent family could be used to expand the prior art list of the topic patent as they disclose the same invention.

\section{Data \& Evaluation Tracks}
\label{patent-retrieval-data}
This section presents an overview of evaluation tracks organized for patent data analysis along with available data collections with focus on tasks pertaining to PR.

\subsection{CLEF-IP Collections}
The Conference and Labs of the Evaluation Forum\footnote{http://www.clef-initiative.eu/} (CLEF) is an European series of workshops which started in 2001 to foster research in Cross Language Information Retrieval (CLIR). The Intellectual Property (IP) track (CLEF-IP) which ran between (2009-2013) was organized to: 1) foster research in patent data analysis, and 2) provide large and clean test collections of multi-language patent documents, specifically in the three main European languages (English, French, and German). Research labs have the opportunity to test their methods on multiple shared tasks such as PR, patent classification, image-based PR, image classification, flowchart recognition, and structure recognition (\cite{roda2010,piroi2010clef,piroi2011clef,piroi2012,piroi2013}).

The CLEF-IP data collection are patent documents extracted from USPTO, EPO and WIPO data. It is provided through the Information Research Facility\footnote{http://www.ir-facility.org/} (IRF) and hosted by Marec \footnote{http://www.ifs.tuwien.ac.at/imp/marec.shtml}. Patent documents are provided in XML format and have common Document Type Definition (DTD) schema. The collection was constructed according to the proposed methodology by \cite{graf2008methodology} and is divided into two pools: 
\begin{enumerate}
	\item \textbf{The corpus pool}: documents selected from this pool are provided for participating labs as training or lookup instances.
	\item \textbf{The topics pool}: documents selected from this pool are called topics and they represent testing or evaluation instances. For example, in prior art search, the topic might be a patent application document for which it is required to retrieve prior art. In this scenario, the patents that constitute prior art are called relevance assessments and obtained from the corpus pool.
\end{enumerate}

The XML documents consist of the main textual sections such as {\it bibliographic data}, {\it abstract}, {\it description}, and {\it claims}. Each section is written in one or more languages (English, French, and/or German) and is denoted by a language code. At least the claims of granted patents (B1 documents) are written in the three languages because it is EPO requirement once a patent application is granted. 

\textbf{CLEF-IP 2009 Collection:}
this dataset was designed for the prior art search task \cite{roda2010}. The corpus pool contains documents published between (1985-2000) ($\sim$2m documents pertaining to $\sim$1m unique patents). The topics pool contains documents published between (2001-2006) ($\sim$0.7m documents pertaining to $\sim$0.5m individual patents). Topics are sets of documents from the topics pool with sizes ranging from 500 to 10,000 topics. Topics were assembled from granted patent documents including {\it abstract}, {\it description}, and {\it claims} sections. Citation information from the {\it bibliographic data} section was excluded.

A major pitfall in this dataset is its topics, which were chosen from granted patent documents (B1 documents). Initially, the creators of the dataset were motivated by having topics from granted patent documents which have claims in three languages. This was thought to provide a kind of parallel corpus suitable for CLIR. The problem of using such documents is simple, it contradicts the practice of IP search professionals who start with the patent application document not the granted one.

\textbf{CLEF-IP 2010 Collection:}
this dataset was created for the prior art search and patent classification tasks \cite{piroi2010clef}. The corpus pool of this dataset contains documents with publication date before 2002 ($\sim$2.6m documents pertaining to $\sim$1.9m unique patents). The topics pool contains documents published between (2002-2009) ($\sim$0.8m documents pertaining to $\sim$0.6m unique patents). Topics for the prior art task are two sets of documents from the topics pool; a small set of 500 topics and a larger set of 2000 topics. Unlike the CLEF-IP 2009 dataset, topics are assembled from patent application documents rather than granted patent documents. 

\textbf{CLEF-IP 2011 Collection:}
This dataset was created as a test collection for four tasks: prior art search, patent classification, image-based prior art search, and image classification \cite{piroi2011clef}. The topics and corpus pools were the same as in CLEF-IP 2010 dataset. For the prior art task, 3973 topics were provided as a separate archive of patent application documents. 

\textbf{CLEF-IP 2012 Collection:}
this dataset was created as a test collection for three tasks: passage retrieval starting from claims, chemical structure recognition, and flowchart recognition \cite{piroi2012}. The topics and corpus pools were the same as in CLEF-IP 2010 dataset. The passage retrieval task is designed differently from previous CLEF-IP prior art search collections. The purpose for this tasks is to retrieve both documents and passages relevant to a set of claims. Topics for the passage retrieval task were extracted from patent applications published after 2001. Relevance judgments were the highly relevant citations only (i.e., marked X or Y) in the examiners' search reports (A4 documents) of chosen topic patents.

\textbf{CLEF-IP 2013 Collection:}
this dataset was created as a test collection for two tasks: 1) passage retrieval from claims, and 2) structure recognition from patent images \cite{piroi2013}. The topics and corpus pools were the same as in CLEF-IP 2010 dataset. Similar to CLEF-IP 2012, the CLM task is designed to retrieve both documents and passages relevant to a set of claims. Topics for the passage retrieval task were extracted from patent applications published after 2002. Overall, the topics set contained 148 topics extracted from 69 patent applications.

\subsection{NTCIR Collections}
The Japanese National Institute of Informatics Testbeds and Community for Information access Research project\footnote{http://research.nii.ac.jp/ntcir} (NTCIR) started in 1997 to support research in IR and other areas, focusing on CLIR. NTCIR has been organizing a series of workshops providing test collections to researchers for evaluating their methodologies on multiple CLIR tasks \cite{ntcircolltbl}. Between NTCIR-3 and NTCIR-11 (2002-2013), there has been dedicated tasks for patent data analysis including patent retrieval \cite{iwayama2003overview}, classification, mining, and translation. 

\textbf{NTCIR-3:}
the PR task in NTCIR-3 targeted the "technology survey" problem. The dataset for this task includes: 1) full text of Japanese patent applications between (1998-1999), 2) {\it abstract} of Japanese patent applications between (1995-1999) along with their respective English translations, and 3) 30 search topics where each topic includes a related newspaper article. The task is to retrieve patents relevant to news articles. Both cross-genre experiments in which patents were retrieved by a newspaper clip as well as ordinary ad hoc retrieval of patents by topics were conducted \cite{iwayama2003overview}.

\textbf{NTCIR-4:}
two PR tasks were organized in NTCIR-4 \cite{fujii2004overview}: 1) patent map generation, and 2) invalidity search. The dataset for the PR tasks includes: 1) unexamined Japanese patent applications published between (1993-1997) along with English translations of the {\it abstract}, and 2) 34 search topics where each topic is a claim of a rejected patent application which was invalidated because of existing prior art. Relevance judgments were individual patents that can invalidate a topic claim by its own or in conjunction with other patents. Relevant passages to the invalidated claim were also annotated and added to the relevance judgments.

\textbf{NTCIR-5:}
two PR tasks were organized in NTCIR-5 \cite{Fujii05overviewof}: 1) document retrieval (invalidity search), and 2) patent passage retrieval. The dataset for the invalidity search task includes: 1) unexamined Japanese patent applications published between (1993-2002) along with English translations of the {\it abstract}, and 2) 1200 search topics where each topic is a claim of an invalidated patent application. Relevance judgments were generated in a manner similar to the one used in NTCIR-4 invalidity search task.

\textbf{NTCIR-6:}
two PR tasks were organized in NTCIR-6 \cite{fujii2007overview}: 1) Japanese retrieval (invalidity search), and 2) English retrieval. The dataset for the Japanese retrieval task is the same one used in NTCIR-5 but more topics were used (1685 topics). The English retrieval task was focusing on finding all the citations cited by the applicant and the examiner. The dataset for this tasks includes: 1) granted patents from the USPTO between (1993-2000), and 2) 3221 search topics where each topic is a granted patent published between (2000-2001).

\subsection{TREC-CHEM Collections}
The TREC-CHEM track was organized to motivate large scale research on chemical datasets, especially chemical patent retrieval \cite{lupu2009trec}.

\textbf{TREC-CHEM 2009:}
this collection was created as a test collection for two tasks \cite{lupu2009trec}: 1) technology survey, and 2) prior art search. 18 topics were provided for the technology survey task where relevance judgments were obtained from experts and chemistry graduate students. For the prior art search, 1,000 patents were provided as test topics where relevance judgments were collected from the citations of topic patents as well as their family members. The search corpus contains $\sim$1.2m chemical patents filed until 2007 at EPO, USPTO, and WIPO. It also contains 59K scientific articles.

\textbf{TREC-CHEM 2010:}
this collection was created for the same two tasks as in TREC-CHEM 2009 \cite{lupu2010trec}. 30 topics were provided for the technology survey task. The search corpus contains $\sim$1.3m chemical patents and 177K scientific articles. Relevance judgments were created the same way as in TREC-CHEM 2009.

\textbf{TREC-CHEM 2011:} 
this collection was created for the same two tasks as in previous TREC-CHEM tracks besides a new chemical image recognition task. The technology survey task topics were biomedical and pharmaceutical patents \cite{lupu2011overview}.

\subsection{Other Sources}
Other IP data sources are detailed by \cite{schwartz2015data}. These include full patent texts as well as bibliographic information from major patent offices such as the USPTO, EPO, and WIPO. Bibliographic information for patents published from 1976 to 2006 is provided through the National Bureau of Economic Research\footnote{https://sites.google.com/site/patentdataproject/} (NBER) and subsequently cleaned and extended to include patents until 2013\footnote{http://rosencrantz.berkeley.edu/batchsql/}. Patent prosecution histories are available through the Patent Application Information Retrieval\footnote{http://portal.uspto.gov/pair/PublicPair} (PAIR). Patent assignments, filings, classifications, and petition decisions are also provided through the USPTO bulk downloads previously hosted by Google\footnote{https://www.google.com/googlebooks/uspto-patents.html} and now by the USPTO\footnote{https://www.uspto.gov/learning-and-resources/bulk-data-products}.

\section{Patent Retrieval Tasks}
\label{patent-retrieval-tasks}
The goal of PR is to retrieve relevant patent documents to a given search request (query). This request can take different forms such as a sequence of keywords, a memo, or a complete text document (e.g. a patent application). The purpose of this task is manifold, for example:
\begin{itemize}
	\item Retrieve related patents to a given patent application in order to gather related work, or invalidate one or more of its claims.
	\item Explore patent filing activity under specific technology.
	\item Explore the competitive landscape of a given company by looking at other companies filing patents similar to the given company patents.
\end{itemize}

Because of these multiple objectives, various PR tasks were proposed to fulfill each objective, and multiple datasets were provided depending on the given task.

Prior-art search is the main theme of the CLEF-IP and NTCIR tracks. The importance of this task stems from the requirement by all patent offices that filed patents must constitute novel, non-obvious, and non-abstract ideas. Therefore, an important activity through the patent life-cycle is to thoroughly ensure that no earlier published patent or material describing the prescribed ideas exist. The task can be defined as follows:

\textit{Problem: given a patent application X, retrieve all related documents to X.}	

\begin{table*}
	\footnotesize
	\setlength{\tabcolsep}{1.8pt}
	{\renewcommand{\arraystretch}{1.4}
		\centering
		\caption{Scenarios of Patent Prior Art Search}
		\label{tbl:prior-art-scenarios}
		\begin{tabular}{|l|l|l|l|l|}
			\hline
			\multicolumn{1}{|c|}{\textbf{Search Task}} & \multicolumn{1}{c|}{\textbf{Who}} & \multicolumn{1}{c|}{\textbf{When}} & \multicolumn{1}{c|}{\textbf{Purpose}}  & \multicolumn{1}{c|}{\textbf{Output}} \\ \hline
			Related Work                       & \begin{tabular}{@{}l@{}}  inventor/\\prosecutor   \end{tabular}   & pre-grant         & all related work  &  applicant's disclosure                    \\ \hline
			Patentability     & \begin{tabular}{@{}l@{}} prosecutor/\\examiner \end{tabular}   & \begin{tabular}{@{}l@{}} pre-grant/\\examination \end{tabular}          & novelty breaking work     & grant/modify/reject            \\ \hline
			Infringement                   & \begin{tabular}{@{}l@{}} owner/\\investor \end{tabular}       & post-grant                & \begin{tabular}{@{}l@{}} relevant claims/\\infringing products  \end{tabular}   & sue/license/clearance            \\ \hline
			Freedom to Operate                   & investor        & post-grant                & \begin{tabular}{@{}l@{}} relevant claims/\\related work  \end{tabular}   & clearance            \\ \hline
			Invalidity                   & \begin{tabular}{@{}l@{}} competitor/\\defendant \end{tabular}        & post-grant                & novelty breaking work     & \begin{tabular}{@{}l@{}} re-examine/\\inter-parts review/\\post-grant review \end{tabular}            \\ \hline
			Technology Survey                   & \begin{tabular}{@{}l@{}} technology \\analyst \end{tabular}       & pre/post-grant                & all published patents     & survey report            \\ \hline
		\end{tabular}
	}
\end{table*}

Prior art search is a total recall task\footnote{It is required to achieve 100\% recall at acceptable precision.}, therefore it demonstrates several challenges. Search coverage is one of the main challenges, because it is required to cover all previously published material (patent or non-patent literature) in all forms (electronic or printed) which is infeasible. Another major challenge is the need to search through materials written in different languages. Last but not least, traditional IR methods perform poorly when confronted with the patent prior art search task. Mainly because the patent language is full of jargon and user defined terminology. Inventors intentionally tend to use different vocabulary to express same or similar ideas in order to establish the novelty of their work.

Prior art search is performed at different stages of the patent life-cycle, by different stakeholders, for various purposes, and for limited period of time. Understanding the real-life practices of patent professionals is critical to better satisfy their information need \cite{jurgens2012going}. In other words, the search scenario depends on when it is done, by who, and for what reason(s). Table \ref{tbl:prior-art-scenarios} shows these various scenarios which are detailed below.

\textbf{Related Work Search:}
during the pre-grant stage, inventors and prosecutors run related work search to retrieve all relevant work to the invention. Moreover, some patent offices request from inventors an applicant's disclosure document specifying all related publications when filing a new application.

\textbf{Patentability Search:}
during the examination stage, patent examiners perform patentability search in order to ensure that the proposed ideas are novel, non-obvious, and non-abstract. The output of this task would be a search report with all retrieved relevant publications. In this report, each entry will have a special code indicating whether it is just a related publication, or novelty breaking one. Examiners would also specify which passages or figures in retrieved publications constitute relevancy. Depending on the search findings, the patent office might grant, reject, or ask the applicant to modify the patent application. Patentability search is also performed by patent prosecutors as a sanity check. Although this task should be of equal interest to prosecutors who file the patent application as it is to examiners, prosecutors often do not dig deep searching for relevant publications, and delegate finding relevant prior work to examiners in order to save costs.

\textbf{Infringement Search:}
this task, also called product clearance search, aims to ensure whether an existing or a proposed product is infringing any published patent claim(s). Patent owners require that type search to find out if a third party has a product with features that are within the scope of one or more claims of their patents. If so, they might either sue or negotiate a license with that infringing party. 

Investors and R\&D managers, on the other hand, require that type of search to ensure newly proposed product(s) are not infringing a published patent claim(s) and investment in such products would be lucrative. The scope of search in this case would be limited to patent and copyrighted literature only. Deep understanding and correct interpretation of patent claims is imperative for building the correct correspondence between product features and claims in order to establish or dismiss infringement.

\textbf{Freedom to Operate Search:}
this PR task extends beyond infringement search. Here, investors and R\&D managers not only need to make sure that proposed products do not infringe an existing patent or copyrighted material, but also to ensure they have the freedom to file patents on these products without worrying about previously prior art that might invalidate such inventions. Another objective of freedom to operate search is to make better investment decisions and R\&D plans according to existing prior art.

\textbf{Invalidity Search:}
as patents guarantee monopoly rights to their owners on the economic value of granted inventions, companies and other parties usually monitor granted patents of their competitors or pertaining to their technology landscape to ensure competitive superiority. Therefore, invalidity search is performed to find published material that was missed by the patent office during patentability search. Invalidity search is also considered as the first line of defense when a party is confronted with patent infringement lawsuit. Again published material might include patent or non-patent literature such as books, news articles, academic periodicals...etc. After finding such validity breaking material, a third party might file a post-grant (opposition) procedure depending on the patent office policies. For example, the USPTO provides procedures such as re-examination, inter-partes Review\footnote{http://www.uspto.gov/patents-application-process/appealing-patent-decisions/trials/inter-partes-review}, and post-grant review\footnote{http://www.uspto.gov/patents-application-process/appealing-patent-decisions/trials/post-grant-review} in front of the Patent Trial and Appeal Board\footnote{https://ptabtrials.uspto.gov} (PTAB).

\textbf{Technology Survey:}
another PR task where, in a typical scenario, business managers would request search professionals to prepare a survey of patent documents given a memorandum they prepared from some source (e.g., news article). In the PR task at NTCIR-3 (\cite{iwayama2003overview}), this basic scenario was limited to patent literature and it was assumed that patent documents are just a collection of technical papers. 

\begin{figure*}
	\centering
	\includegraphics[scale=0.5]{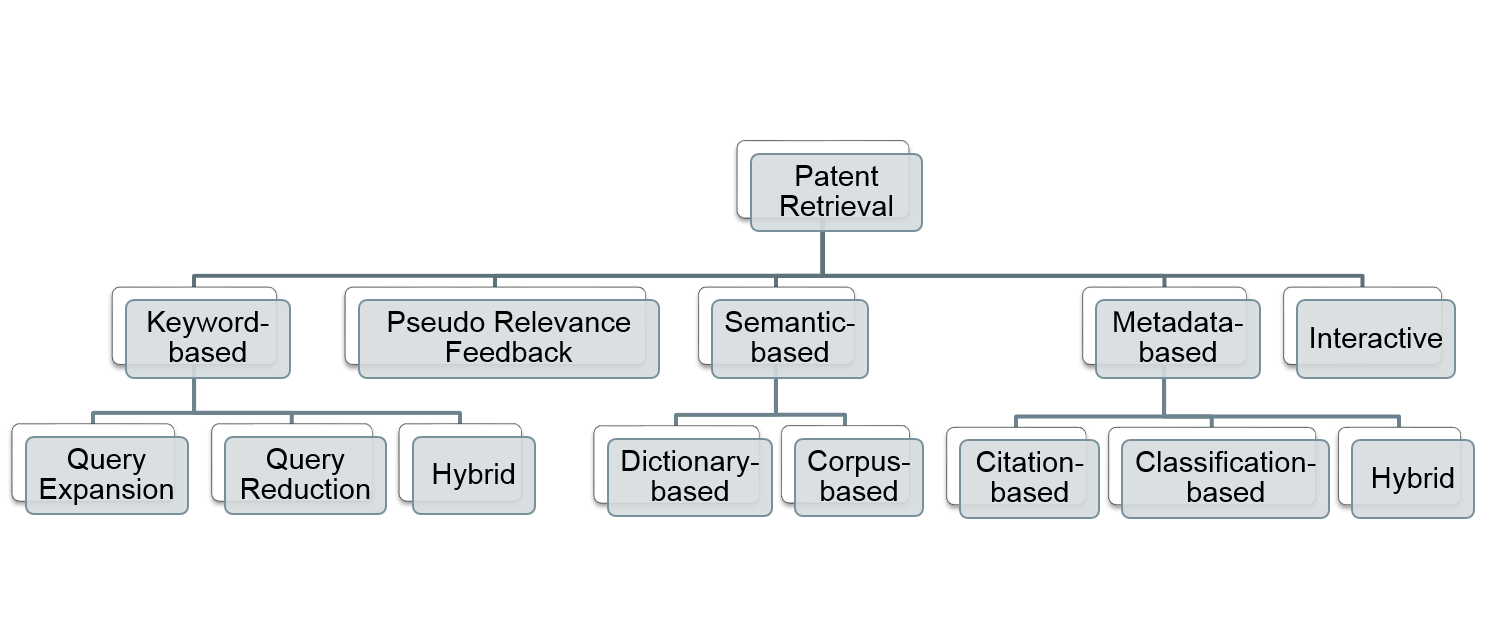}
	\caption{Taxonomy of Patent Retrieval Methods}
	\label{fig:pr-methods}
\end{figure*}

\section{Patent Retrieval Methods}
\label{patent-retrieval-methods}
In this section, we present a comprehensive review of PR methods and approaches. We start by presenting available test collections and evaluation metrics. Then, we provide a taxonomy of these approaches highlighting their characteristics and limitations.

As shown in Figure \ref{fig:pr-methods}, PR methods can be categorized depending on which piece(s) of data from both the search queries and the search corpus are used for retrieving relevant documents. Keyword-based methods utilize only terms from search queries and look for exact matches in the target corpus. Pseudo Relevance Feedback methods utilize terms from the top ranked results of running the initial query to improve the set of relevant retrieved results. Semantic-based methods try to overcome the vocabulary mismatch problem between the search terms and related patents vocabulary by matching them based on their meanings. Metadata-based methods exploit the language independent non-textual metadata and bibliographic information in order to improve patent retrievability. Finally, interactive methods aim to better organize and present search results to the users. Moreover, through interaction, users are engaged in an iterative process of searching, reviewing, and refining hoping to retrieve as many relevant results as possible.

\subsection{Test Collections \& Evaluation Measures}
As we highlighted in section \ref{patent-retrieval-data}, several datasets were created to support evaluating different PR techniques. In almost all of these datasets, relevant documents to search queries were collected from the citations of topic patent documents (e.g., CLEF-IP 2009/2010/2011 collections). Because these citations represent related prior work, they are appropriate only for the related work search task.

In other datasets such as CLEF-IP 2012/2013 collections, relevant documents were collected from novelty breaking citations found in examiners' search reports, therefore, these datasets are appropriate for the patentability and invalidity search tasks. Though invalidity search requires non-patent literature as well.

Standard information retrieval as well as patent retrieval specific evaluation measures are generally used to evaluate patent retrieval systems including: 
\begin{enumerate}
	\item Precision ({\it P}) and Recall ({\it R}) at top-{\it K} ranks (e.g., {\it K}=\{1, 5, 10, 50, 100, 1000\}). 
	\item Mean Average Precision (MAP) \cite{baeza1999modern} which generally favors early retrieval of relevant documents with less focus on recall.
	\item Normalized Discounted Cumulative Gain (nDCG) \cite{jarvelin2002cumulated} which favors not only early retrieval of relevant documents but also the respective ranking quality of these documents.
	\item Patent Retrieval Evaluation Score (PRES) \cite{magdy2010pres} which was proposed specifically for recall-oriented tasks such as PR. PRES focuses on the overall system recall as well as user's review effort which can be estimated from the rankings at which relevant documents are retrieved.
\end{enumerate}

\begin{table}
	{\renewcommand{\arraystretch}{1.4}
		\scriptsize
		\setlength{\tabcolsep}{1.8pt}
		\centering
		\caption{Keyword-based Patent Retrieval Methods}
		\label{tbl:kw-retrieval}
		\begin{tabular}{|p{0.25\textwidth}|p{0.4\textwidth}|l|c|c|c|c|}
			\hline
			\multicolumn{1}{|c|}{\textbf{Method}} & \multicolumn{1}{c|}{\textbf{Description}} & \multicolumn{1}{c|}{\textbf{Dataset}} & \multicolumn{1}{c|}{\textbf{MAP}}  & \multicolumn{1}{c|}{\textbf{P}}  & \multicolumn{1}{c|}{\textbf{R}}  & \multicolumn{1}{c|}{\textbf{PRES}} \\ \hline		
			\vspace{-4.5mm}
			\cite{verberne2009prior} & 
			\vspace{-5.5mm}
			\begin{itemize}[noitemsep,leftmargin=1.1em]
				\item remove stopwords, and punctuation
				\item use claims as BOW
				\vspace{-2mm}
			\end{itemize}		
			& \begin{tabular}{@{}l@{}} clef-ip \\2009 \end{tabular}  & 0.05 & 0.01 & 0.22 & \multicolumn{1}{c|}{-}  \\ \hline
			\vspace{-4.5mm}
			\cite{magdy2009exploring} & 
			\vspace{-5.5mm}
			\begin{itemize}[noitemsep,leftmargin=1.1em]
				\item remove stopwords, and frequent terms
				\item use different sections with manual weights
				\item perform IPC filtering
				\item use bigrams with tf$>$1		 	
				\vspace{-2mm}
			\end{itemize}		
			& \begin{tabular}{@{}l@{}} clef-ip \\2009 \end{tabular} & 0.12 & \multicolumn{1}{c|}{-} & 0.63 & \multicolumn{1}{c|}{-}  \\ \hline
			\vspace{-4.5mm}
			\cite{mahdabi2011building}$^\star$ & 
			\vspace{-5.5mm}
			\begin{itemize}[noitemsep,leftmargin=1.1em]
				\item use query language models on different sections
				\item use queries of 100 terms
				\item perform IPC filtering
				\vspace{-2mm}
			\end{itemize}		
			& \begin{tabular}{@{}l@{}} clef-ip \\2010 \end{tabular}  & 0.12 & \multicolumn{1}{c|}{-} & 0.60 & 0.49  \\ \hline
			\vspace{-4.5mm}
			\cite{wang2015query}$^\star$ & 
			\vspace{-5.5mm}
			\begin{itemize}[noitemsep,leftmargin=1.1em]
				\item use linguistic-based concepts
				\item concept weighting using weighted tf-idf and mutual information
				\vspace{-2mm}
			\end{itemize}		
			& \begin{tabular}{@{}l@{}} clef-ip \\2010 \end{tabular}  & 0.10 & \multicolumn{1}{c|}{-} & 0.48 & 0.40 \\ \hline
			\cite{konishi2005query} & 
			\vspace{-3.5mm}
			\begin{itemize}[noitemsep,leftmargin=1.1em]
				\item patterns to identify {\it claim} components terms
				\item patterns for explanation terms from {\it description}
				\item rank boosting based on IPC  
				\vspace{-2mm}
			\end{itemize}		
			& ntcir-5 & 0.20 & \multicolumn{1}{c|}{-} & \multicolumn{1}{c|}{-} & \multicolumn{1}{c|}{-}  \\ \hline
			\multicolumn{7}{l}{$\star$ indicates scores @1000} \\
		\end{tabular}
	}
\end{table}

\subsection{Query Reformulation (QRE)}
The most widely used techniques for patent retrieval are the Query Reformulation (QRE) techniques. These methods aim at transforming the input query $Q$ into $\bar{Q}$ by means of reduction or expansion of $Q$ terms in order to improve the retrievability of relevant documents. QRE can be performed through: 
\begin{itemize}
	\item \textbf{Query Reduction (QR):} where a representative subset of terms are selected from $Q$ and used as $\bar{Q}$ terms. Position-based methods are the most commonly used in this category where terms from specific parts or sections of the patent document are used, or given higher matching weight than others. Another example of query reduction is the IPC-based methods which utilize terms from IPC definitions as a lexicon or stop-words list for $Q$.
	\item \textbf{Query Expansion (QE):} where representative terms other than the ones in $Q$ are extracted and merged with $Q$ to form $\bar{Q}$. Pseudo Relevance Feedback (PRF) methods are the most prominent in this category where terms from top ranked results of running $Q$ are used to expand $Q$ terms assuming these top results are relevant \cite{cao2008selecting}. Other semantic-based query expansion methods work by expanding $Q$ with terms of similar meanings such as synonyms or hyponyms.
	\item \textbf{Hybrid (query expansion \& reduction):} where irrelevant terms are removed from $Q$ and more relevant terms are appended to $Q$ to form $\bar{Q}$. Most techniques used for query expansion are appropriate for query reduction as well, where only terms appearing in the expansion list are kept and all others are pruned.
\end{itemize}

\subsubsection{Keyword-based Methods}
This set of techniques retrieves relevant documents by looking for exact matches between search query term(s) and the target data. keyword search operates under the closed vocabulary assumption where vocabulary is derived solely from terms that appear in the target search data. Table \ref{tbl:kw-retrieval} shows some keyword-based methods along with their performance results on benchmark datasets. Keyword-based techniques differ in: 1) which elements of the target data are indexed, 2) which query terms are selected/removed, 3) the relative weights of such terms, and 4) the match scoring function.

\textbf{Query Reduction (QR):} the rationale behind QR approaches is intuitive as patents are very long documents with several sections. Querying with the whole document would be impractical and inefficient. Some query reduction methods are position-based; they select relevant terms based on their position in the patent document \cite{verberne2009prior,d2010clef,wanagiri2010prior,magdy2009exploring,mahdabi2011building}. For example, \cite{verberne2009prior} used only terms from the {\it claims} section on the CLEF-IP 2009 collection. However, the results were moderate in terms in MAP compared to other runs on the same collection. 

\cite{magdy2009exploring} experimented using text from different sections of the topic patent on the CLEF-IP 2009 collection. The authors used various combinations of sections including: 1) short sections such as {\it title, abstract}, first line of the {\it description}, first sentence of the {\it claims}, and 2) lengthy sections such as the {\it description} and the {\it claims}. The authors assigned different weights to each section manually. Their best scores were achieved using a combination of all short sections and post-filtering retrieved documents keeping only those that share the same IPC classification code with the topic patent. The main challenge with such approach is how to assign the respective weight of each section automatically. Moreover, IPC filtering wouldn't be possible when only partial patent application is available for prior art search.

\cite{mahdabi2011building} proposed a position-based query reduction method which selects relevant query terms by building two query language models using various sections of the topic patent: 1) a variant of the weighted log-likelihood model \cite{meij2009query} , and 2) a model based on the parsimonious language model \cite{hiemstra2004parsimonious}. Their experiments showed that queries constructed from terms in the {\it description} section using weighted log-likelihood give better results than other sections which agrees with the previous results (\cite{xue2009transforming, magdy2010applying, bouadjenek2015study}). The main advantage of this approach is that, respective weights of query terms are derived automatically from the query model. However, some challenges still exist regarding tuning the model parameters such as the smoothing parameter which was set heuristically.

\begin{table*}
	{\renewcommand{\arraystretch}{1.4}
		\scriptsize\setlength{\tabcolsep}{1.8pt}
		\centering
		\caption{Pseudo Relevance Feedback Patent Retrieval Methods}
		\label{tbl:prf-retrieval}
		\begin{tabular}{|p{0.3\textwidth}|p{0.4\textwidth}|p{0.13\textwidth}|p{0.07\textwidth}|p{0.07\textwidth}|}
			\hline
			\multicolumn{1}{|c|}{\textbf{Method}} & \multicolumn{1}{c|}{\textbf{Description}} & \multicolumn{1}{c|}{\textbf{Dataset}} & \multicolumn{1}{c|}{\textbf{MAP}}  & \multicolumn{1}{c|}{\textbf{PRES}} \\ \hline		
			\cite{bouadjenek2015study} & 
			\vspace{-4mm}
			\begin{itemize}[noitemsep,leftmargin=1.1em]
				\item use different methods of query expansion and reduction from the PRF set
				\item use Rocchio, MMR, LM
				\vspace{-2mm}
			\end{itemize}		
			& 
			\vspace{-4mm}
			\begin{itemize}[noitemsep,leftmargin=0.5em,label={}]
				\item clef-ip 2010
				\item clef-ip 2011
				\vspace{-2mm}
			\end{itemize}		
			& 
			\vspace{-4mm}
			\begin{itemize}[noitemsep,leftmargin=0.5em,label={}]
				\item 0.13
				\item 0.10
				\vspace{-2mm}
			\end{itemize}		
			& 
			\vspace{-4mm}
			\begin{itemize}[noitemsep,leftmargin=0.5em,label={}]
				\item 0.55
				\item 0.45
				\vspace{-2mm}
			\end{itemize} \\ \hline
			\cite{magdy2009exploring} & 
			\vspace{-4mm}
			\begin{itemize}[noitemsep,leftmargin=1.1em]
				\item naive PRF
				\item remove stop-words
				\item use most frequent terms
				\vspace{-2mm}
			\end{itemize}		
			& \multicolumn{1}{c|}{clef-ip 2009}	& \multicolumn{1}{c|}{0.05} & \multicolumn{1}{c|}{-} \\ \hline
			\cite{mahdabi2012learning} & 
			\vspace{-4mm}
			\begin{itemize}[noitemsep,leftmargin=1.1em]
				\item build regression model using relevance score, RF similarities...etc
				\item use the model to estimate the effectiveness of RF
				\item use top 100 RF and maximize AP
				\vspace{-2mm}
			\end{itemize} 
			&  \multicolumn{1}{c|}{clef-ip 2010}	&  \multicolumn{1}{c|}{0.16} & \multicolumn{1}{c|}{0.56} \\ \hline
			\cite{ganguly2011patent} & 
			\vspace{-4mm}
			\begin{itemize}[noitemsep,leftmargin=1.1em]
				\item perform query segmentation
				\item retain segments highly to be generated using RF LM
				\vspace{-2mm}
			\end{itemize} 
			& \multicolumn{1}{c|}{clef-ip 2010}	& \multicolumn{1}{c|}{0.14} & \multicolumn{1}{c|}{0.47} \\ \hline
			\cite{golestan2015term} & 
			\vspace{-4mm}
			\begin{itemize}[noitemsep,leftmargin=1.1em]
				\item manually annotate one relevant RF result
				\item add terms in the annotated result to the query
				\vspace{-2mm}
			\end{itemize} 
			&  \multicolumn{1}{c|}{clef-ip 2010}	& \multicolumn{1}{c|}{0.29\footnotemark} & \multicolumn{1}{c|}{-} \\ \hline				
			\cite{golestan2015term} & 
			\vspace{-4mm}
			\begin{itemize}[noitemsep,leftmargin=1.1em]
				\item assume relevant RF results are known
				\item add terms more frequent in relevant than irrelevant RF to query
				\vspace{-2mm}
			\end{itemize} 
			& \multicolumn{1}{c|}{clef-ip 2010}	& \multicolumn{1}{c|}{0.48\footnotemark} & \multicolumn{1}{c|}{-} \\ \hline
		\end{tabular}
	}
\end{table*}

\textbf{Query Expansion (QE):} pattern-based QE was proposed in many studies (\cite{osborn1997evaluating,wang2015query,konishi2005query}). \cite{wang2015query} proposed patterns in the form of syntactic rules in order to extract query terms as weighted concepts. \cite{konishi2005query} proposed a pattern-based query expansion method for the patent invalidity search task on the NTCIR-5 collection. In this task the initial query of the topic patent was the terms in the {\it claims} section. However, rather than using only raw {\it claims} terms which are often abstract, \cite{konishi2005query}, using pattern matching, identifies other specific terms in the {\it description} and use them as expansion terms. First, components of the invention are extracted from the topic {\it claim} using handcrafted patterns. Secondly, explanation sentences describing components of the invention are extracted from the {\it description} using handcrafted patterns. Thirdly, terms from first and second steps are used as the new query. The results showed that this query expansion approach works better than using terms extracted from the {\it claims} section only. The main drawback of this method is its dependency on manually coded patterns to identify potential terms. Meanwhile, it demonstrates the potential of using entities and their relations as retrieval features motivating the need for deeper and more generic linguistic analysis of patent texts.

\subsubsection{Pseudo Relevance Feedback (PRF)} 
These methods are one of the prominent techniques used for QRE. PRF starts with an initial run of the given query $Q$. Then, terms from top ranked results are used to select, remove, and/or expand terms in $Q$, assuming that these top results are relevant. PRF is thus advantageous as it works automatically without human intervention but might be computationally inefficient especially with long queries. Table \ref{tbl:prf-retrieval} shows some PRF methods along with their performance results on benchmark datasets.

\footnotetext[17]{This is a semi-supervised performance}
\footnotetext[18]{This is an Oracle performance}

Despite their effectiveness and popularity, several challenges arise when it comes to PRF-based QRE \cite{bouadjenek2015study} such as: 1) which part(s) of the patent application should be used as the initial query?; 2) which part(s) of the retrieved results should be used as the source of expansion and/or reduction?; 3) what is the best length of the expansion list in case of query expansion, or the best threshold for removing terms in case of reduction?; 4) which pseudo-relevant results are really relevant and how many of them should be used?; and 5) what is the best relevance scoring model for the search task (e.g., BM25 \cite{robertson1995okapi}, the vector space model with tf-idf weighting...etc).

\cite{bouadjenek2015study} provided a thorough evaluation on the CLEF-IP 2010/2011 collections to address some of the above challenges. The authors explored the scenario when only partial patent application is available for prior art search (e.g., {\it title, abstract, extended abstract}, or {\it description}). The authors tested different query expansion and reduction general methods such as Rocchio \cite{Salton:1971:SRS:1102022} and a variant of the Maximal Marginal Relevance (MMR) \cite{carbonell1998use}. They also tested patent-specific methods utilizing synonym sets \cite{magdy2011study}, language models \cite{ganguly2011patent}, and IPC-based lexicon \cite{mahdabi2013leveraging}. After experimenting various sections as sources for the initial query terms as well as expansion/reduction sources, the results showed that, the {\it description} section among other sections is the best to use as the initial query in case of both query expansion and reduction. query reduction was not beneficial for the long {\it description} queries as it already contains good coverage of relevant terms. However, query reduction on {\it description} queries was useful as it removed many of the noisy terms. Generally, query reduction outperformed query expansion on {\it description} and {\it extended abstract} queries which indicates that, with long queries, query reduction is effective for better retrieval performance. The results also showed that generic query expansion methods such as Rocchio works generally better for query expansion than patent-specific query expansion methods. Finally, the results showed that BM25 scoring works better than the TF-IDF scoring on the long {\it description} queries for both query reduction and expansion, while TF-IDF works better than BM25 on short and medium-length {\it title} or {\it abstract} queries. Through this comprehensive experimental study, the authors did not evaluate the impact of using multiple sections in combination as sources for query expansion or reduction. More importantly, the study does not provide any insights about the respective values of number of expansion terms or term removal threshold and whether these values are somewhat deterministic or vary widely calling for interactive setting.

To address the problem of poor PRF results in patent retrieval compared to traditional information retrieval, \cite{bashir2010improving} proposed a novel approach for PRF-based query expansion which builds a model that learns to identify better PRF results based on their similarity with the query patent over specific terms. These terms are learned by building a classification model that classifies whether a term would be useful for query expansion or not according to some proximity features between the original query terms and pseudo-relevant terms. The authors, through experiments on a subset of USPTO patents, showed the ability of this model to introduce more relevant query expansion terms and subsequently increasing the retrievability of individual patents. However, the authors did not evaluate this model on a any of the available test collections. Moreover, extracting similarity features and computing similarities with PRF results during query execution is computationally expensive and time consuming. 

Along the same efforts, \cite{mahdabi2012learning} proposed a framework for identifying effective PRF documents at runtime and then performing query expansion using terms from these relevant documents. The authors first proposed patent-specific features and then used them to build a regression model which calculates a relevancy score of each PRF document. Though results on the CLEF-IP 2010 collection were encouraging, several challenges still exist. For example, the computational complexity of calculating the regression model features at runtime. And PRF parameters tuning (e.g., number of PRF documents to use).

\cite{ganguly2011patent} proposed a PRF approach which utilizes a language model for query reduction of long queries composed of full patent applications. The authors argued that, naive application of PRF to expand query terms could add noisy terms causing query-topic drift. Moreover, naive removal of terms that has unit term frequency in the query could cause removal of useful terms and thus hurt retrieval effectiveness. Instead, the authors proposed a PRF-based query reduction technique which generates language model similarity scores between query segments (sentences or n-grams) and top ranked results. Segments with top scores are kept and all others are removed. Results on the English subset of the CLEF-IP 2010 collection showed that the proposed approach outperforms the baselines. Parameter tuning is still the main downside of this technique. The performance of the proposed approach was unstable compared to the baselines with different parameter values. Specifically, the window size, the number of pseudo-relevant documents, and the fraction of terms to retain.

\cite{golestan2015term} provided a study on hybrid QRE which aims to automatically approximate the optimal $\bar{Q}$ by careful selection/expansion of relevant query terms. To motivate the efficacy of QRE on retrieval performance, the authors first designed an experiment where relevance judgments of a query patent $Q$ were assumed to be known in advance. After running $Q$, using PRF on top-{\it k} documents, only terms that are more frequent in retrieved relevant documents (those from relevance judgments) than irrelevant documents are kept and used as $\bar{Q}$. Then, querying using $\bar{Q}$ achieved a better performance than state-of-the-art on the English subset of CLEF-IP 2010 collection. To approximate $\bar{Q}$ automatically, the authors proposed four different methods hoping to identify relevant vs. irrelevant terms in $Q$ by: 1) removing terms with high document frequency in the top-100 retrieved documents, 2) removing infrequent terms in $Q$, 3) using frequent terms in relevant documents assuming the top-5 retrieved documents are relevant, and 4) performing query reduction on $Q$ using IPC definitions as stop-words. All of the four methods failed to perform better than the keyword-based baseline. More interestingly, the authors demonstrated that, baseline performance can be doubled if only one relevant document was manually provided by the user. This last observation motivates the need for interactive QRE as a simple and effective method for patent retrieval.

\begin{table*}
	{\renewcommand{\arraystretch}{1.4}
		\scriptsize\setlength{\tabcolsep}{1.8pt}
		\centering
		\caption{Semantic-based Patent Retrieval Methods}
		\label{tbl:semantic-retrieval}
		\begin{tabular}{|l|p{0.4\textwidth}|p{0.12\textwidth}|p{0.07\textwidth}|p{0.07\textwidth}|}
			\hline
			\multicolumn{1}{|c|}{\textbf{Method}} & \multicolumn{1}{c|}{\textbf{Description}} & \multicolumn{1}{c|}{\textbf{Dataset}} & \multicolumn{1}{c|}{\textbf{MAP}}  & \multicolumn{1}{c|}{\textbf{PRES}} \\ \hline		
			\cite{magdy2011study} & 
			\vspace{-4mm}
			\begin{itemize}[noitemsep,leftmargin=1.1em]
				\item use Wordnet synonyms and hyponyms for query expansion
				\item slow processing time
				\item no improvement
				\vspace{-2mm}
			\end{itemize}		
			& clef-ip 2010 & 
			\vspace{-4mm}
			\begin{itemize}[noitemsep,leftmargin=0.3em,label={}]
				\item 0.136
				\item 0.140\textsuperscript{$\star$}
				\vspace{-2mm}
			\end{itemize}		
			& \vspace{-4mm}
			\begin{itemize}[noitemsep,leftmargin=0.3em,label={}]
				\item 0.484
				\item 0.486\textsuperscript{$\star$}
				\vspace{-2mm}
			\end{itemize} \\ \hline
			\begin{tabular}{@{}l@{}}
				\cite{tannebaum2012acquiring} \\ 
				\cite{tannebaum2012analyzing} \\ 
				\cite{tannebaum2013mining} \\
				\cite{tannebaum2014using} \\
				\cite{tannebaum2015patnet} \\
				\cite{tannebaum2015effect}
			\end{tabular} &
			\vspace{-13.10mm}
			\begin{itemize}[noitemsep,leftmargin=1.1em]
				\item mine query logs for synonyms, co-occurring, and proximity terms
				\item no improvement
				\item use upon request
				\vspace{-3mm}
			\end{itemize} 
			& \vspace{-12.5mm}
			clef-ip 2010	& 
			\vspace{-13.7mm}
			\begin{itemize}[noitemsep,leftmargin=0.3em,label={}]
				\item 0.139
				\item 0.139\textsuperscript{$\star$}
				\vspace{-3mm}
			\end{itemize} & 
			\vspace{-13.7mm}
			\begin{itemize}[noitemsep,leftmargin=0.3em,label={}]
				\item 0.512
				\item 0.512\textsuperscript{$\star$}
				\vspace{-3mm}
			\end{itemize} \\ \hline
			\cite{magdy2011study} & 
			\vspace{-4mm}			
			\begin{itemize}[noitemsep,leftmargin=1.1em]
				\item using synonyms learned from parallel translations (EN, GE, and FR)
				\item improve MAP only
				\item use upon request
				\vspace{-3mm}
			\end{itemize}		
			& clef-ip 2010	& 
			\vspace{-4mm}			
			\begin{itemize}[noitemsep,leftmargin=0.3em,label={}]
				\item 0.144
				\item 0.140\textsuperscript{$\star$}
				\vspace{-2mm}
			\end{itemize}		
			& 
			\vspace{-4mm}
			\begin{itemize}[noitemsep,leftmargin=0.3em,label={}]
				\item 0.485
				\item 0.486\textsuperscript{$\star$}
				\vspace{-2mm}
			\end{itemize}	 \\ \hline		
			\multicolumn{5}{l}{$\star$ indicates baseline performance}
		\end{tabular}
	}
\end{table*}

\subsubsection{Semantic-based Methods}
As we mentioned before, in PR queries can vary from few terms (e.g., survey memo) to thousands of terms (e.g., full patent application). Straightforward keyword-based PR proved to be ineffective simply because of the vocabulary mismatch between query terms and relevant patents content. \cite{magdy2009exploring} showed that, in the CLEF-IP 2009 collection, 12\% of the relevant documents have no common words with the search topics. This motivates the need for novel approaches to bridge this vocabulary mismatch gap. Several semantic-based methods have been proposed in attempt to match queries with relevant documents based on their meanings rather than relying on keyword matches only.  Table \ref{tbl:semantic-retrieval} shows some semantic-based methods along with their performance results on benchmark datasets.

\textbf{Dictionary-based:} 
semantic-based methods perform QRE by expanding the query to include other terms that have similar meanings to the original query terms. The first category of these methods are the dictionary-based techniques which use either generic \cite{magdy2011study}, technical \cite{lopez2009multiple}, or patent-specific dictionaries \cite{tannebaum2012analyzing,tannebaum2013mining,tannebaum2015patnet,tannebaum2015effect,Wajda2016,mahdabi2014patent} for QRE. Generic dictionaries could be existing lexical databases such as WordNet \cite{wordnet}, while patent-specific dictionaries are lexical databases generated from patent-related data such as examiner's query logs. In either case, similar or related terms to the original query terms are retrieved from such dictionaries and used for query expansion.

\cite{magdy2011study} explored the use of WordNet for query expansion in patent retrieval on the CLEF-IP 2010 collection. Overall, adding synonyms and hyponyms for nouns and verbs in the original query increased the MAP score slightly, while decreased the PRES score significantly. Moreover, query execution time was increased considerably. The authors considered this a "negative" result. As the use of WordNet was proven to be effective in other retrieval tasks \cite{voorhees1998using,liu2004effective}, more experiments are needed to affirm the authors' conclusion. For example, investigating the impact of using synonyms only or hyponyms only, and expanding terms belonging to specific sections or ambiguous terms only.

Recently, more research was focused on utilizing domain-specific and technical dictionaries rather than WordNet. Examiners' query logs have been an important resource for building such technical thesauri. \cite{tannebaum2012acquiring,tannebaum2012analyzing,tannebaum2013mining,tannebaum2014using,tannebaum2015patnet} and \cite{tannebaum2015effect} introduced an analysis of the USPTO examiners' search query logs. Their analysis, though on a subset of query logs, revealed interesting insights about patent examiners' search behavior which could be very useful for designing effective patent retrieval systems. For example, the authors noted that about examiners' behavior while searching for prior art: 1) the average query length is four terms, 2) search terms are mostly from the patent application under investigation, 3) expansion terms represent small percentage of query terms and mostly appear in the specific patent domain terminology, 4) the majority of query terms represents subject technical features that appears in the {\it claims} section, while very little percentage of them appears in the {\it description} section, 5) the majority of terms are nouns, followed by verbs, then adjectives, and 6) about half of the query operators used are "OR", followed by "AND", then proximity operators. 

Tannebaum et al. built upon these insights and introduced methods to automatically identify synonyms/equivalents, co-occurring terms, and proximity relations for expanding query terms by mining examiners' search logs. As we can notice, learning expansion terms from query logs might be misleading because not all query sessions succeed to identify prior art. Additionally, deeper analysis of the query logs considering other metadata such as relevant hits count might be useful in this regard. On the other hand, it would be more useful if we can model the features of these terms, for example, based on their location, frequency, part-of-speech...etc. From effectiveness perspective, evaluating the generated lexical knowledge on the CLEF-IP 2010 collection did not record significant improvement \cite{tannebaum2015effect}. Therefor, the authors recommended using it in an interactive mode rather than automatic mode to semi-automate query generation.

\begin{table*}
	{\renewcommand{\arraystretch}{1.4}
		\setlength{\tabcolsep}{1.8pt}
		\centering
		\caption{Metadata-based Patent Retrieval Methods}
		\label{tbl:metadata-retrieval}	
		\begin{tabular}{|p{0.25\textwidth}|p{0.4\textwidth}|p{0.13\textwidth}|p{0.07\textwidth}|p{0.07\textwidth}|}
			\hline
			\multicolumn{1}{|c|}{\textbf{Method}} & \multicolumn{1}{c|}{\textbf{Description}} & \multicolumn{1}{c|}{\textbf{Dataset}} & \multicolumn{1}{c|}{\textbf{MAP}}  & \multicolumn{1}{c|}{\textbf{PRES}} \\ \hline		
			\cite{fujii2007enhancing} & 
			\vspace{-4.10mm}
			\begin{itemize}[noitemsep,leftmargin=1.1em]
				\item use PageRank on patents citation graph
				\item use patent popularity among top results with weighted voting
				\vspace{-2mm}
			\end{itemize}		
			& ntcir-6 & 
			\vspace{-4.10mm}
			\begin{itemize}[noitemsep,leftmargin=0.3em,label={}]
				\item 0.075
				\item 0.081
				\item 0.071\textsuperscript{$\star$}
				\vspace{-7mm}
			\end{itemize}
			& \multicolumn{1}{c|}{-} \\ \hline
			\cite{mahdabi2014effect} & 
			\vspace{-4.10mm}
			\begin{itemize}[noitemsep,leftmargin=1.1em]
				\item build query specific citation graph from PRF results and their citations
				\item weight nodes using PageRank
				\item estimate query LM from the graph nodes considering their PageRank scores
				\vspace{-2mm}
			\end{itemize}		
			& clef-ip 2011	& 
			\vspace{-4.10mm}
			\begin{itemize}[noitemsep,leftmargin=0.3em,label={}]
				\item 0.105
				\item 0.099\textsuperscript{$\star$}
				\vspace{-2mm}
			\end{itemize}		
			& 
			\vspace{-4.10mm}
			\begin{itemize}[noitemsep,leftmargin=0.3em,label={}]
				\item 0.481
				\item 0.450\textsuperscript{$\star$}
				\vspace{-2mm}
			\end{itemize}	 \\ \hline
			\cite{mahdabi2014query} & 
			\vspace{-4.10mm}
			\begin{itemize}[noitemsep,leftmargin=1.1em]
				\item using time-aware random walk on weighted citation graph
			\end{itemize} 
			& clef-ip 2011	& 
			\vspace{-4.10mm}
			\begin{itemize}[noitemsep,leftmargin=0.3em,label={}]
				\item 0.125
				\item 0.058\textsuperscript{$\star$}
			\end{itemize} & 
			\vspace{-4.10mm}
			\begin{itemize}[noitemsep,leftmargin=0.3em,label={}]
				\item 0.536
			\end{itemize} \\ \hline				
			\multicolumn{5}{l}{$\star$ indicates baseline performance}
		\end{tabular}
	}
\end{table*}

\textbf{Corpus-based:} 
the second category of semantic-based QRE is the corpus-based methods. In these methods, textual corpora are analyzed to extract semantically related concepts to query terms which can be used for query expansion. \cite{al2014wikipedia} proposed a Wikipedia-based query expansion method which works by first creating a summary of each Wikipedia article containing the main category, all titles under the main category, and other categories with in/out links to the main category. At query time, query terms and phrases are matched with page summaries, then, phrases from matching pages are scored and selected for query expansion under the assumption that they are semantically related. Experiments on the subset of USPTO patents in the NTCIR-6 collection showed increase in MAP over other query expansion techniques. However, the authors used IPC codes rather than citations as relevance judgments to topic queries which does not reflect the typical search practices, where it is needed to retrieve related patent documents not related classification codes.

Another corpus-based method was proposed by \cite{magdy2011study}, where synonym sets were automatically generated from the CLEF-IP patent corpus. The authors utilized parallel translations of patent sections in different languages to build a word-to-word translation model and infer synonymy relation when a word in one language is translated to multiple words in another language. These multiple words under some probabilistic threshold could be considered synonyms. Overall results using this method were better than PRF and Wordnet based query expansion, but worse than the keyword-based baseline in \cite{magdy2010applying}. The authors also showed that, the performance of this method on some topics was better than the baseline which indicates its potential. The issue they raised is how to more effectively apply query expansion by selecting "good" terms \cite{cao2008selecting}, or predicting query expansion performance beforehand \cite{cronen2002predicting,lv2009adaptive}. Such challenges can also be alleviated semi-automatically by developing intelligent and usable interactive query expansion frameworks which engage users in such decision. Finally, \cite{krestel2013recommending} applied topic modeling of search hits in order to better rank retrieved patents. The results on a small collection of the USPTO patents showed improved MAP.

\subsubsection{Metadata-based Methods}
Patents are not only textual documents, they contain lot of non-textual metadata and bibliographic information as well (e.g., citations, tables, formulas, drawings, classification...etc). Combining metadata analysis with text-based PR has shown improvements in performance in the literature \cite{fujii2007enhancing,lopez2009multiple,lopez2010experiments,eisinger2013automated,mahdabi2014effect}. Metadata features are also language independent making them advantageous when used for CLIR. Table \ref{tbl:metadata-retrieval} shows some metadata-based methods along with their performance results on benchmark datasets

\textbf{Citation-based:} The use of citation analysis for better patent retrieval is the most heavily reported technique of metadata-based methods. Naively incorporating citations from topic patent applications as prior art proved to be effective, eliminating the need for deeper citation analysis (\cite{magdy2011simple}. However, citation extraction from patent texts is challenging because there is no standard writing style for patent references. \cite{lopez2010experiments} developed a tool for citation mining which identifies, parses, normalizes, and consolidates patent citations. As citations might not be always available in all scenarios (e.g., related work search, technology survey...etc), more mature techniques are needed. \cite{fujii2007enhancing} proposed using PageRank \cite{brin2012reprint} and document popularity as an additional scoring to re-rank query top results returned using {\it claims}-based queries. The results of applying popularity scoring on the English subset of NTCIR-6 improved MAP and recall over the raw text-based scoring. Incorporating PageRank, though intuitive, poses many challenges especially because patent documents have references to non-patent literature which would produce incomplete citation graph. \cite{mahdabi2014effect} extended their query modeling technique in  \cite{mahdabi2011building} by incorporating term distributions of the PRF results as well as their citations in calculating the query language model. The authors first construct a query-specific citation graph using PRF results and their citations and assign a score for each of them using PageRank. Then, a query model is estimated from term distributions of the documents in the citation graph constrained by their respective PageRank. Finally, query expansion is performed using the estimated query model. Experiments on the CLEF-IP 2011 collection showed improved recall performance with no change in precision, which indicates the usefulness of using cited documents vocabulary for query expansion. Best improvements were achieved using the top 30 PRF documents, 2-levels citation graph, and 100 expansion terms. However, we can notice two main computational challenges using this technique in real-time setting: 1) computing the PageRank of the 2-level citations graph, and 2) estimating the query model from top PRF documents as well as documents in the citation graph.

\textbf{Classification-based:} 
these methods utilize classification information of the topic patent and the retrieved documents to improve the performance of patent retrieval \cite{kim2006cluster,harris2009role,harris2010comparison,chen2011ipc,giachanou2015multilayer}. The naive use of IPC classification is to filter retrieved documents to keep only ones that share the same IPC classification code at some level (e.g., same subclass) with the topic patent \cite{magdy2009exploring,gobeill2009simple}. More sophisticated use of classification information was introduced by \cite{verma2011patent} who proposed a new representation of patent documents based on IPC classifications. The method utilizes IPC codes assigned to the corpus patents as well as codes of their citing documents to form an IPC class vector. First, the vector is initialized from patent's IPC code, then codes of citing patents are propagated over multiple iterations. The most similar patents are retrieved using cosine similarity between IPC class vectors and re-ranked using text-based search utilizing the top 20 tf-idf topic patent terms. Experiments on the CLEF-IP 2011 collection showed improved recall but low MAP scores. The instability of the patent classification system poses a real challenge when it comes to incorporating classification metadata into PR systems. Overtime, new classes are added to the classification hierarchy and existing classes are expanded. In order to do reliable search based on classification codes, these changes must be accounted for periodically. Moreover, patents are assigned to multiple classification codes, however, almost all previous research considered only the primary class but not secondary classifications which might, if utilized, improve the retrieval performance.

textbf{Hybrid:} these methods utilize various sources of metadata to improve PR performance. \citet{mahdabi2014query} built upon previous work in \cite{mahdabi2011building} and \cite{mahdabi2014effect} and proposed a query expansion method that utilizes time-aware random walk on a weighted patent citations network. Citation weights are derived from various metadata (e.g., classification codes, inventors, assignee...etc). Citations with higher weights are considered more influential when performing query expansion. Experiments on the CLEF-IP 2010/2011 collections show improved recall and MAP. \citet{mahdabi2014patent} proposed building a query-specific lexicon from IPC definition pages and using it for query expansion. Unfortunately, the lexicon would be helpful only if the query represents a complete patent document with IPC codes assigned to it which is not always the case especially at the early stages of the patent life-cycle.

\subsubsection{Interactive Methods}
Interactive patent retrieval is inevitable. As we can notice from the above review, effective fully automated retrieval of patent prior art is very challenging. Best methods perform around average in terms of PRES and much less in terms of MAP. Additionally, these methods require tuning a large number of parameters and thresholds whose optimal values differ according to the given query and the specific information need. For example, deciding which patent section to use?, which PRF results?, and which expansion terms and their respective weights?. The answers of these questions are not deterministic and probably require multiple interaction cycles with the user in order to satisfy his/her information need.

\begin{figure}
	\centering
	\includegraphics[scale=0.5]{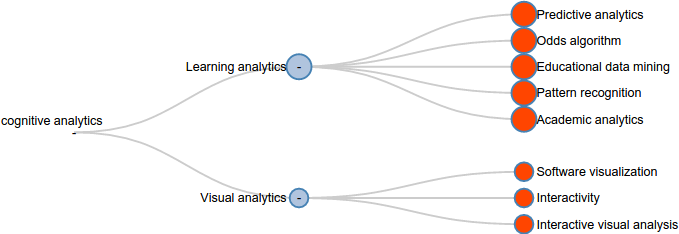}
	\caption{Concept graph of Cognitive Analytics. Light blue nodes are explicit concepts and red nodes are latent concepts.}
	\label{fig:msa-visual_landscape}
\end{figure}

\begin{figure}
	\centering
	\includegraphics[scale=0.5]{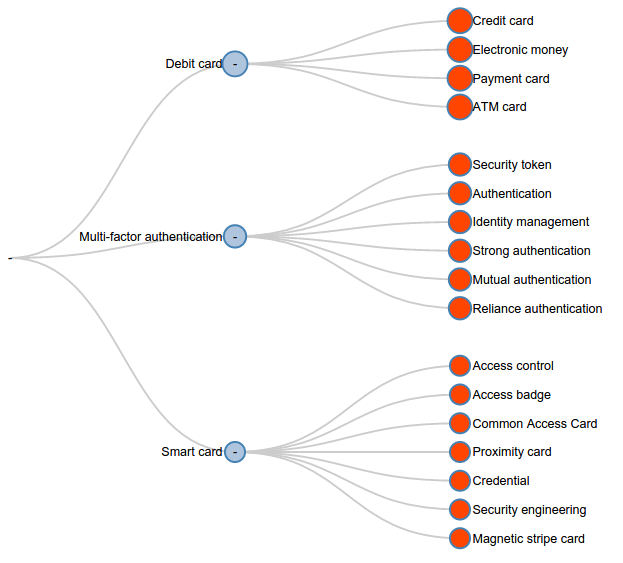}
	\caption{Concept graph using Bank of America's 100 patent titles. Light blue nodes are explicit concepts and red nodes are latent concepts.}
	\label{fig:msa-visual_bofa_fig}
\end{figure}

Current interactive methods in patent retrieval are more focused on better organization, integration, and utilization of structured and textual patent data than on better retrieval performance. In other words, patent retrieval is addressed as a professional search problem rather than prior art search problem. \citet{fafalios2014exploratory} presented a keyword-based interactive search framework to support patent search. The interaction elements are presented through post-analysis of search results in the form of facets based features like static metadata (e.g., IPC codes), textual clustering, named entity extraction, semantic enrichments, and others. The framework was applied on patent search \cite{salampasis2014perfedpat} and evaluated using user study of twelve patent examiners \cite{salampasis2014evaluation}. Evaluation responses indicated overall acceptance of the framework in terms of usability, ease of use, efficiency, learnability. However, the authors did not report on the effectiveness or success of the system helping patent examiners to find prior art. 

\cite{shalabyinnomsaflairs2016} proposed a visual interactive semantic framework for patent analysis which features semantic-based query expansion of search queries using Mined Semantic Analysis (MSA) \cite{shalaby2015measuring}. In a nutshell, MSA builds an association knowledge graph using rule mining of concept rich textual corpora (e.g., Wikipedia). After mining the "See Also" link graph of Wikipedia, MSA could represent a topic query as a Bag-of-Concepts (BOC) derived from the association knowledge graph. This BOC could then be used to expand the original query terms. Figure \ref{fig:msa-visual_landscape} shows an example of the query expansion map of {\it Cognitive Analytics}. Another example is presented in Figure \ref{fig:msa-visual_bofa_fig} showing concept map of 10 patents of {\it Bank of America} using the {\it abstract} section. Users can interact with the concept map by removing nodes and updating the search results. \cite{shalaby2016visual} demonstrated the applicability of their framework to support tasks such as prior art search, competitive intelligence, technology landscape analysis and exploration. However, they did not provide a controlled study evaluating the performance of their method on benchmark collections.

Developing interactive methods for patent retrieval is also motivated by recent analysis which showed significant performance improvement if only one relevant document was manually provided by the user \cite{golestan2015term}. Performance gains using Technology Assisted Review (TAR) \cite{grossman2011technology,cormack2014evaluation} in domains like electronic discovery motivates investigating the applicability of machine learning TAR protocols in patent retrieval.

Technology assisted review, like patent retrieval, is a total-recall task where it is required to find all relevant documents to the search request with reasonable effort (time and cost). It is thus a human-in-the-loop process where a human expert manually annotates a subset of the documents as relevant or irrelevant. The underlying algorithm subsequently builds a ranking model by training on such annotations and uses this model to promote more relevant results and demote irrelevant ones as more documents are searched and annotated. This process stops when enough results are obtained. Typically, these algorithms utilize techniques such as continuous active learning combined with Boolean search in order to develop and adapt the ranking model \cite{grossman2011technology}. 

Several questions still need to be addressed when it comes to investigating technology assisted review protocols applicability to patent retrieval, as these protocols were only evaluated in ad hoc search scenarios. The complexity of patents terminology and availability of multiple sources of metadata would, likely, demonstrate many opportunities for adaptation and modifications to the current technology assisted review protocols.

\section{Related Topics}
\label{patent-retrieval-related}
Despite intense interest within the research community in patent retrieval, the patent industry has many other challenges and open problems which are of high interest and value to various stakeholders, such as economists, R\&D managers, and legal professionals, to name a few. In this section, we try to lightly touch on these tasks and highlight some challenges and possible future directions.

\subsection{Patent Quality Assessment}
Assessing the technical quality and importance of inventions is very important to patent owners because it allows them to: 
\begin{itemize}
	\item better utilize their IP management costs by automated recommendation of patent maintenance decisions.
	\item better determine the novelty and originality of their patents.
	\item maximize licensing revenues by automatic estimation of the patent value.
\end{itemize}
Because there is no ground truth for quality measurements, performance evaluation of quality assessment techniques is usually based on indicators such as correlation with patent forward citations, maintenance status history, court rulings (if any), and/or patent reexamination history (if any). Some early work scored patents using their metadata such as citations count, maintenance history, global prosecution efforts \cite{lanjouw1998count}, and even manually by patent attorneys. Automated patent quality assessment has gained more traction in recent years though.

Citation analysis has been and still a main technique for patent valuation \cite{trajtenberg1990penny,harhoff1999citation,hall2005market,czarnitzki2010market,wang2014exploring}. \cite{wang2014exploring} proposed a probabilistic mixture approach to predict whether a topic patent will be renewed at different renewal periods. The method first divides the citations into two groups; technological and legal. From each group, different features reflecting the technological richness, technological influence, legal patent scope, and legal blocking power of each patent are combined. The authors subsequently build a binary classifier using these probabilistic features. Evaluation is performed by comparing the model's predictions against the renewal decisions of a collection of patents. While proved effectiveness, estimating patent value as a binary outcome might not be practical especially if a patent owner needs to prioritize his maintenance decisions of multiple patents. 

Quality assessment based on the lexical features of the patent text was also explored in the literature \cite{jin2011patent,liu2011latent,hido2012modeling}. \cite{liu2011latent} proposed a graphical model to estimate patent quality as a latent variable. The model utilized lexical features extracted from the patent text such as {\it claims} n-grams age and popularity, lexical alignment between the {\it claims} and the {\it description}, number of dependent and independent claims, number of reported classes when filing the patent, and other features. The authors also incorporated measurements such as forward citations count, court decisions, and reexamination records. It is clear that court decisions are only available for small number of patents which might not allow building a robust model. 

\cite{jin2011patent} modeled the patent maintenance decision as recommendation problem where patents were represented as multimodal heterogeneous information network. The model utilized several metadata features, lexical features such as unique words and lengths of different sections, as well as inventor and assignee profile features. Experimental results showed high prediction accuracy on a large number of USPTO patents. 

\cite{hu2012finding} proposed a time-based topic model which ranks patents novelty and influence based on whether the dominant topics in patent's prior art (for novelty) or forward art (for influence) are still active topics. The authors also proposed using time decay function to address the problem of old patents having less prior art and more forward art than newer patents and vice versa. Results showed high correlation between assigned ranks and forward citations count. 

\cite{hido2012modeling} proposed a scoring model which assigned a patentability score to each patent and thus can be utilized to determine whether it will be granted. First, the authors extracted textual features such as word frequency, word age, and syntactic complexity (e.g., number of sentences). Then, they trained a classifier using previous patent office decisions as ground truth. Though results showed the model effectiveness, the utilized syntactic complexity features are all extracted from the topic patent and thus could be good predictors for the writing quality not patentability potential.

The correlation between patent {\it claims} novelty and patent value using lexical analysis of patent text has been analyzed in previous studies (\cite{chen2009simple,hasan2009coa}). \cite{hasan2009coa} proposed an IR-based ranking tool which analyses patent {\it claims} for originality. The technique first extracts key terms and phrases from the {\it claims} text using syntactic patterns and then looks for usage patterns backward to determine their novelty, and forward to determine their influence. The method considers usage patterns only through user defined time window. It is also keyword-based and hence will fail to capture key phrases that are semantically similar and subsequently might give inaccurate scores.

Along the efforts of using patent legal data for quality assessment, \cite{mann2012new} utilized prosecution histories, court decisions, and patent textual features to analyze patent quality. The analysis suggested that patent examination records would be very helpful in better discriminating high from low quality patents and possibly improve the examination process as a whole.

\subsection{Patent Litigation}
Litigation in general, and patent litigation specifically has been and still a topic of interest to legal professionals. With the increased amounts of digitized data available and the need for technology support in analyzing and mining these huge datasets, litigation became of more interest to computational science researchers. Patent litigation can take many forms, the most common is patent infringement litigation where a patent owner (plaintiff) accuses another party (defendant) of using his/her invention without license or permission. Because litigation is very expensive, the most common defensive action for the defendant is to establish invalidity of the plaintiff invention by issuing a post grant proceeding such as post-grant review or inter-parts review. Now the problem becomes a patent retrieval task, i.e. invalidity search, where one of the aforementioned methods can be utilized with wider scope to cover not only patent literature but also other published material.

The task of automatically establishing patent infringement is not addressed in literature. Such task requires extensive human expertise and reasoning to build correspondences between product features and patent claims. On the other hand, statistical and visual analytics of previous court decisions have shown some degree of success in helping lawyers to better understand possible outcomes and better plan on defense strategies \cite{harbert2013law,allison2014understanding,osbeck2015using}. 

For example, \cite{allison2015our} provided a statistical study on patent cases filed from 2008 to 2009 and decisions made between (2009-2013). The study showed that, there is a strong correlation between court decision, and patent-specific, litigation-specific, and industry-specific variables such as industry and technology type, inventors foreign status, number of claims, number of forward and backward citations, and number of defendants sued. 

\cite{rajshekhar2016analytics} studied the potential of concept-based semantic search in patent litigation. The authors designed an experiment in order to retrieve invalidating patents to a given litigated patent using a subset of PTAB's final decisions as ground truth and a search corpus of $\sim$7m USPTO patents. The authors, based on the experimental results and through interviews with patent practitioners, concluded that, a one-size-fits-all semantic search approach is incapable of capturing the highly nuanced relevance
judgments made in the domain of patent litigation. Rather, the search workflow should be modeled as a multistage information seeking process, where users are presented with interactive elements to control the search space, and their feedback is incorporated iteratively in the relevance ranking of retrieved results for enhanced performance. 

Finally, There is much to be done in building predictive models for patent litigation given the availability of prior case datasets that were not available few years ago (e.g., prosecution histories, court decisions, and PTAB decisions).

\subsection{Technology Licensing}
Patents represent one of the most valuable assets in today's enterprises which, if leveraged effectively, guarantee not only competitive superiority, but also huge licensing revenues \cite{chen2009simple}. The technology licensing task is three sided. First, patent owners would be interested in finding potential licensees with reasonable effort. Second, licensees would like to find relevant inventions to their businesses. Third, owners and businesses would be interested in gauging the strategic and protection values of a patent in order to support their pricing and offering decisions. 

While there is no much research focusing on automatically recommending potential licensees, the task of recommending patents to be licensed was relatively more considered. \cite{chen2009simple} proposed a platform called SIMPLE which is used at IBM to identify target patents for licensing. Given a set of topic patents, SIMPLE uses nearest neighbor similarity to find other patents that are most similar to the given topic set. Then, all the patents are grouped and proposed as one licensing package to interested party. The platform was extended in \cite{spangler2010simple} to allow retrieving target patents using free text search. We can notice that current trends for identifying potential patents for licensing model the problem as a PR task. More elaboration on the SIMPLE platform was introduced by \cite{spangler2011exploratory} using interactive visualization. First, portfolios of two companies are contrasted to find content overlap between both of them using proximal search. Then, the closest patents to the overlap area are recommended as candidates for licensing. 

\section{Concluding Remarks}
\label{patent-retrieval-conclusion}
In this paper we presented a comprehensive review of patent retrieval methods and approaches. It is clear that the well-performing information retrieval techniques in areas like Web search cannot be utilized directly in PR without deliberate domain adaptation and customization. Furthermore, state-of-the-art performance in automatic patent retrieval is still low ($<$0.2 MAP). Several proposed techniques for query expansion, query reduction and pseudo relevance feedback require tuning of various parameters. Search professional practices suggest that effective prior art search requires multiple iterations of searching, reviewing, and refining. On the other hand, examiners' query formulation practices  (few keywords and Boolean search) are different from those of automatic methods (many keywords and free-text search). These observations motivate the need for interactive search tools which provide cognitive assistance to search professionals with minimal effort. These tools must also be developed in hand with patent professionals considering their practices and expectations.

Unexplored patent-related data sources might be an opportunity for breakthrough improvements over the current modest state-of-the-art in patent retrieval. For example, utilizing reexamination records, PTAB decisions, differences between the patent application and the granted version, examiner/applicant correspondences, and prosecution histories. All these resources are not yet fully explored in the literature of patent retrieval.

Related tasks such as patent quality assessment, litigation, and licensing are of less focus among computational scientists. However, they provide wide opportunities for future exploration from computational and modeling perspectives. These tasks require interdisciplinary and cooperative efforts from both legal professionals and the computer science research community.

\bibliographystyle{spbasic}
\bibliography{references}   
\end{document}